\documentclass[a4paper,fleqn]{cas-sc}
\usepackage[numbers]{natbib}
\usepackage{bytefield}
\usepackage{rotating}
\usepackage{booktabs}
\usepackage{color}
\usepackage{wasysym}
\usepackage{amssymb}
\usepackage{etoolbox}
\usepackage{alltt}
\usepackage{hyperref}
\usepackage{float}

\usepackage{caption}
\usepackage{subcaption}
\usepackage{multirow}
\usepackage{enumerate}
\usepackage{listings}

\usepackage{textcomp}

\usepackage{changepage}
\usepackage{url}
\PassOptionsToPackage{hyphens}{url}\usepackage{hyperref}
\usepackage{hyperref}

\usepackage{amsthm}
\usepackage{amsmath,amssymb,amsfonts}
\usepackage{url}

\usepackage{xcolor}
\usepackage{comment}
\usepackage[shortlabels]{enumitem}
\usepackage[utf8x]{inputenc} %Linux
\usepackage{array}

\DeclareCaptionLabelFormat{andtable}{#1~#2  \&  \tablename~\thetable}

\usepackage{placeins}
\usepackage{balance}
\usepackage{epsfig}

\usepackage{color}
\usepackage{calc}
\usepackage{amstext}
\usepackage{graphicx}
\graphicspath{{../}}
\usepackage{makecell}

\usepackage{algorithm,algpseudocode}
\usepackage{multicol}
\usepackage{url}
\usepackage[underline=true]{pgf-umlsd}
\usetikzlibrary{positioning,calc}
\usetikzlibrary{decorations.pathreplacing}
\usepackage{placeins}

\usepackage{xparse}%

\usepackage{blindtext}

\usepackage{morewrites}

\usepackage{wrapfig}
\usepackage{xkeyval}%

\newwrite\authorbibfile%

\AtBeginDocument{%
	\immediate\openout\authorbibfile=\jobname.aub%
}%

\AtEndDocument{%
	\immediate\closeout\authorbibfile
	\InputIfFileExists{\jobname.aub}{}{}
}%

\makeatletter

\define@key{authorbib}{scale}[1]{%
	\def\AuthorbibKVMacroScale{#1}%
}

\define@key{authorbib}{wraplines}[10]{%
	\def\AuthorbibKVMacroWraplines{#1}%
}

\define@key{authorbib}{imagewidth}[4cm]{%
	\def\AuthorbibKVMacroImagewidth{#1}%
}

\define@key{authorbib}{overhang}[10pt]{%
	\def\AuthorbibKVMacroOverhang{#1}%
}

\define@key{authorbib}{imagepos}[l]{%
	\def\AuthorbibKVMacroImagepos{#1}%
}

\makeatother

\presetkeys{authorbib}{imagepos=l,imagewidth=4cm,wraplines=15,overhang=20pt}{}

\newlength{\AuthorbibTopSkip}
\newlength{\AuthorbibBottomSkip}
\NewDocumentCommand{\authorbibliography}{+o+m+m+m}{%
	\IfNoValueTF{#1}{%
	}{%
		\setkeys{authorbib}{#1}%
		\immediate\write\authorbibfile{%
			\string\begin{wrapfigure}[\AuthorbibKVMacroWraplines]{\AuthorbibKVMacroImagepos}[\AuthorbibKVMacroOverhang]{\AuthorbibKVMacroImagewidth}^^J
				\string\includegraphics[scale=\AuthorbibKVMacroScale]{#2}^^J
				\string\end{wrapfigure}^^J
		}%
	}%
	\IfNoValueTF{#3}{%
		\typeout{Warning: No author name}%
	}{%
		\immediate\write\authorbibfile{%
			\unexpanded{\vspace{\AuthorbibTopSkip}}^^J
			\string\noindent\relax
			\unexpanded{\textbf{#3}\par}^^J
			\string\noindent\relax
			\unexpanded{#4}^^J%
			\unexpanded{\vspace{\AuthorbibBottomSkip}}^^J
		}%
	}%
}%

\begin{document}
	\let\WriteBookmarks\relax
	\def\floatpagepagefraction{1}
	\def\textpagefraction{.001}
	\shorttitle{PETIoT: PEnetration Testing the Internet of Things}
	\shortauthors{Bella et~al.}
	
	\title [mode = title]{PETIoT: PEnetration Testing the Internet of Things}

	\author[1]{Giampaolo Bella}[orcid=0000-0002-7615-8643]
	\ead{giamp@dmi.unict.it}
		
	\address[1]{Dipartimento di Matematica e Informatica, Universit\`{a} degli Studi di Catania, Catania, Italy}
	
	\author[1]{Pietro Biondi}[orcid=0000-0003-1795-2836]
	\ead{pietro.biondi@phd.unict.it}
	
	\author[2]{Stefano Bognanni}[orcid=0000-0001-5843-2031]
	\ead{stefano.bognanni@leonardo.com}
\address[2]{Cybersecurity Division, Leonardo S.p.A., Catania, Italy}

	\author[3]{Sergio Esposito}[orcid=0000-0001-9904-9821]
	\ead{sergio.esposito.2019@live.rhul.ac.uk}
\address[3]{Department of Information Security, Royal Holloway University of London, Egham, UK}

\begin{abstract}
Attackers may attempt exploiting Internet of Things (IoT) devices to operate them unduly as well as to gather personal data of the legitimate device owners'. Vulnerability Assessment and Penetration Testing (VAPT) sessions help to verify the effectiveness of the adopted security measures. However, VAPT over IoT devices, namely VAPT targeted at IoT devices, is an open research challenge due to the variety of target technologies and to the creativity it may require.
Therefore, this article aims at guiding penetration testers to conduct VAPT sessions over IoT devices by means of a new cyber Kill Chain (KC) termed PETIoT. \textcolor{black}{Several practical applications of PETIoT confirm that it is general, while its main novelty lies in the combination of attack and defence steps.}
PETIoT is demonstrated on a relevant example, the best-selling IP camera on Amazon Italy, the TAPO C200 by TP-Link, assuming an attacker who sits on the same network as the device's \textcolor{black}{in order to assess all the network interfaces of the device}. Additional knowledge is generated in terms of \textcolor{black}{three zero-day vulnerabilities found and practically exploited on the camera, one of these with High severity and the other two with Medium severity by the CVSS standard}. These are camera Denial of Service (DoS), motion detection breach and video stream breach. The application of PETIoT culminates with the proof-of-concept of a home-made fix, based on an inexpensive Raspberry Pi 4 Model B device, for the last vulnerability. \textcolor{black}{Ultimately, our responsible disclosure with the camera vendor led to the release of a firmware update that fixes all found vulnerabilities, confirming that PetIoT has valid impact in real-world scenarios.} 
\end{abstract}


\maketitle

\section{Introduction}\label{sec:intro}
The Internet of Things (IoT) has revolutionised the current technological landscape. It has changed people's uses of traditional devices and has, at the same time, sparked off the definition of new ones. It would have been otherwise impossible, for example, to operate kitchen appliances from a smartphone and to conceive a smart plug. 
The IoT has virtually no boundaries. Notably, it spans over the application domain of \textit{Cyber-Physical Systems} (CPSs) and that of Information-Operation Technology (IT/OT). While CPSs include layman's devices and applications such as those in the automotive and healthcare areas, IT/OT concerns industrial systems, where manufacturing and production are increasingly intertwining and becoming computer-controlled.

Correspondingly with a rise in pervasiveness, the IoT has experienced a rise in terms of cybersecurity and privacy threats. Once the devices are connected to the Internet, criminals gain a potential opportunity to abuse them and deny their services; for example, the Mirai malware built a botnet to compromise thousands of devices~\cite{mirai}. People's data is often the ultimate target of exploitation due to the fact that IoT devices, particularly CPSs, collect a variety of user preferences and generated data. A car will process data such as cabin preferences and history of visited places and, in healthcare, an ultrasound machine will handle data such as personally identifiable information and health conditions. An additional surge of attacks has been reported during the COVID-19 pandemic~\cite{iotscandal21}.

The need to assess IoT devices from a cybersecurity and privacy standpoint is, therefore, strongly justified, \textcolor{black}{hence the relevance of the contents of this article}. One of the relevant approaches is to instruct a team of ethical hackers to carry out that assessment empirically, often impersonating criminals, albeit in a controlled environment. Such activities stand under the big umbrella of Vulnerability Assessment and Penetration Testing (VAPT), hence an ethical hacker is also addressed as a tester. 
There are various incarnations of VAPT, such as red-blue-purple teaming, white-black-grey approaches and social-engineering assessments, but these are not pivotal for the sequel of this manuscript. VAPT sessions happen repeatedly over IoT devices, for example, at device manufacturing time if the manufacturer follows a Secure Software Development Lifecycle. Alternatively, they may be conducted by potential buyers of the device with the aim of operating a market choice on the basis of the security guarantees that the device offers.

\textcolor{black}{Given the importance of VAPT in the IoT domain, this article poses the following research questions: \textit{How to approach a VAPT session with ethical purposes over IoT devices? Do approaches exist that could be leveraged profitably? And what tailoring would these require?} 
These questions are relevant because VAPT is challenging in general, requiring  creativity beyond sheer competence, but is all the more challenging, in particular, over IoT devices due to the domain size and variety of adopted technologies. In its aim to address the research questions stated above, this article makes the research hypotheses that VAPT is not fully unstructured as a process, and that it is possible to distil out and frame its fundamental steps specifically towards the IoT. This is done through the formalization of a Kill Chain (KC), each KC being a comprehensive description of the various steps forming an attack.}

\subsection{Contributions}

This article demonstrates how to go about VAPT over IoT devices and, at the same time, demonstrates that existing, general approaches are useful but require appreciable rethinking.
\textcolor{black}{This article introduces PETIoT, a new cyber Kill Chain (KC) to conduct VAPT sessions over IoT devices.} PETIoT
%
%It 
is slim, so that testers may begin to apply and experience it quickly.
The value of PETIoT is to enable the penetration tester who approaches the IoT domain with a clearly defined roadmap to conduct her experiments.
As a limitation, it must be stressed that, while a KC is meant to spell out the steps to take to conduct the activities, it does not prescribe the very tools to use through each step and, most importantly, does not teach how to use them. A few basic tools will be suggested below as examples for use during the demonstration of the KC, but it is firm that many alternatives exist and that the tester may use their favourite ones.

By applying PETIoT using basic freeware tools only, we found three vulnerabilities on the TP-Link TAPO C200, the best-seller domestic camera on Amazon Italy at the time of the research. The vulnerabilities discussed in this article can be briefly outlined and evaluated (according to CVSS ~\cite{cvss}) as follows:
\begin{itemize}
\item Vulnerability 1 -- Improper neutralization of inbound packets allows complete Denial of Service. It means that the tester can DoS the camera and make it unusable. This vulnerability has severity 6.5, that is, Medium. We exploited it by an intensive scan through a vulnerability scanner.
\item Vulnerability 2 -- Insufficient entropy in encrypted notifications allows breach of motion detection. It means that the tester can infer whether the camera detects motion despite the fact that the corresponding notifications are encrypted. The tester can also drop the relevant packets, thereby causing a denial of the motion detection service. This vulnerability has severity 5.4, that is, Medium. We exploited it by TLS packet inspection in a dedicated experiment.
\item Vulnerability 3 -- Cleartext transmission of video stream allows breach by unintended actors. It means that the tester can intercept and interpret the full video stream if the camera users run third-party video players.  This vulnerability has severity 8.8, that is, High. We exploited it by traffic eavesdropping in a dedicated experiment.  We also  prototype a fix for this vulnerability by leveraging an inexpensive Raspberry Pi 4 Model B device~\cite{raspb} to transmit the video stream through a cryptographic tunnel.
\end{itemize}

The three vulnerabilities were zero-day vulnerabilities when we found them because they had not been reported before. During the responsible disclosure process, the vendor acknowledged all issues discussed in this article and followed our advice on how to thwart them; the current firmware release addresses all three vulnerabilities.

In summary, this manuscript answers its research questions by making the following contributions:
\begin{itemize}
    \item 
an account on the relevant existing KCs (\S\ref{sec:ckcaccount});
\item an original tailoring of existing KCs into a new KC that combines attack and defence for the first time and is specific for PEnetration Testing the IoT, here termed PETIoT\footnote{The penetration tester's paradox of attacking for the ultimate goal of defending echoes the paradoxical life of Marcel Andr\'e Henri F\'elix Petiot, who was a doctor as well as a serial killer. Hence the name of our Cyber Kill Chain.
} (\S\ref{sec:ourKC});
\item a practical demonstration of PETIoT during a VAPT session on a relevant, best-seller device, the TP-Link TAPO C200 with firmware 1.0.16 (\S\ref{sec:demo});
\item a discovery of three unknown vulnerabilities on the camera (one of High severity and two of Medium severity), their exploitation in an \textcolor{black}{experimental environment}, their responsible disclosure with the vendor, the design of their fixes and a proof of concept of the most complicated of the fixes, which requires a cryptographic tunnel (\S\ref{sec:VA});
\item ultimately, a compilation of lessons learned and recommendations contributing to strengthen the general cybersecurity posture in the IoT domain (\S\ref{sec:lessons}).
\end{itemize}

\subsection{Generality}
PETIoT has taken a long time to shape up as the way it is defined below. It has been gestated over years of experiments with several IoT devices, starting with printers~\cite{overtrustprinter_noblind} and continuing with VoIP phones~\cite{iotjournal_noblind,papervoip_noblind}. It guided VAPT efforts over modern voice personal assistants, notably favouring the discovery of the AvA attack on Amazon Echo Dot~\cite{sergioasiaccs_noblind}. PETIoT was also used over several IP cameras, including Netvue Orb Mini, Netvue Vigil Cam, Ctronics CTIPC-380C-4MP and  TP-Link TAPO C200, with preliminary results made public about the TAPO C200~\cite{iotsms_noblind}. 

This article conceptualises and describes PETIoT fully for the first time. 
To provide a relevant and complete demonstration of the practical application of our KC, 
the article adopts it during a VAPT session against the current best-seller IP camera on Amazon Italy and discusses the complete findings. PETIoT was born by abstraction and generalisation on top of the innumerable vulnerability assessment and penetration testing experiments conducted so far; also, as a KC, its constituent steps are inherently broad. Therefore, PETIoT is general \textit{by design}.

\subsection{Novelty}
The contributions described above are novel at least in two ways. One concerns the very findings on the Tapo C200, which were already mentioned above. Not only were the three vulnerabilities unknown but, remarkably, we shall see how they arise thanks to focus of PETIoT on the device with its network environment rather than solely on the device as a target to intrude into.

The second element of novelty concerns the methodological approach focusing on ethical hacking, so that PETIoT deliberately aims at guiding VAPT activities more than it seeks out to represent known attacks. 
In consequence, our KC distinctively culminates with a step devoted at fixing the vulnerabilities that may have been found and exploited. On one hand, this is innovative in the KC domain, whose focus normally is on the offensive steps and solely on those, as is the case, to just advance an example, with the popular Mitre ATT\&CK~\cite{mitreattack}, while Mitre D3FEND~\cite{mitredefend} is the only relevant KC for the defensive activities. On the other hand, PETIoT combines attack and defence because this is an obvious necessity for the penetration tester, who cannot complete her reports without detailing clear fixing or mitigation activities for the exploited vulnerabilities.

\subsection{Article Structure}
This manuscript continues by presentng the related work (Section~\ref{sec:RW}). It then introduces PETIoT (Section~\ref{sec:ourKC}) and the basics information and toolkit to apply it in practice (Section~\ref{sec:basics}). It demonstrates PETIoT over the TAPO C200 (Section~\ref{sec:demo}), derives lessons learned and recommendations  (Section~\ref{sec:lessons}) and concludes (Section~\ref{sec:concl}).

\section{Related Work}\label{sec:RW}
The related work can be conveniently partitioned into two groups. One concerns the KCs and the other one is about the security aspects of IP cameras.

\subsection{Cyber Kill Chains}\label{sec:ckcaccount}
A historical account on KCs is due to Grant et al.~\cite{Timgrant2013}. However, at the time of this writing, the best established KCs can be summarised as follows:
\begin{itemize}
    \item Lockheed Martin's Cyber Kill Chain~\cite{cyberkillchain} is the pioneering representative. It sets the scene in 2011 with the 7 essential steps of Reconnaissance, Weaponization, Delivery, Exploitation, Installation, Command and Control, Actions on Objectives.
    \item Mitre ATT\&CK~\cite{mitreattack} is a subsequent KC offering a finer-grained description of an attack over 14 steps. Most importantly, it is enhanced with a knowledge base of relevant techniques.
    \item Unified Kill Chain~\cite{unifiedkillchain} is rather modern, dating back to 2017. It spells out 18 steps, grouped as the three essential macro-steps Initial Foothold, Network Propagation, Action on Objectives.
    \item Expanded Cyber Kill Chain~\cite{malone} extends Lockheed Martin's with an Internal Kill Chain and a Target Manipulation Kill Chain. It thus offers finer and finer detail.
    \item Mitre D3FEND~\cite{mitredefend} ``\textit{is at an early stage and is an experimental research project}''. It  aims at expressing the defensive steps against attacks.
\end{itemize}

Over the years, the trend is clear towards specialising the various steps of Lockheed Martin's CKC further, with the result that KCs become more detailed, hence longer. While detail itself may be useful in general, it may turn out an overkill in certain scenarios, sometimes even reaching redundancy, as it can be argued when KCs are spelled out in the IoT domain. 
The reason is an architectural one. Precisely, because the VAPT target is the IoT device, the latter must be made reachable to the tester, otherwise the activities would be aiming at a different target, such as penetrating the network that hosts the device. A \textit{physical reach} means that the device is physically given to the tester, who may then interact by all available device interfaces, such as Ethernet, Wi-Fi, Bluetooth, etc. A \textit{logical reach} means that the device is only made logically accessible to the tester by means of some interface, typically by means of an IP address. In general, a physical reach allows for a broader test, covering all interfaces, while this is harder to reproduce by a logical reach only. Also, a physical reach normally implies a logical reach, but the opposite does not hold.

In consequence, there appears to be some redundancy through the application of Lockheed Martin's CKC to the IoT domain. For example, Reconnaissance as such, which is aimed at identifying and selecting targets, is void because the target is given, hence this step reduces to the sheer gathering of information about the target, for example through network scanning. Also, Weaponization seems redundant because a basic toolkit may suffice to tackle innumerable devices, as we shall see below. \textcolor{black}{We shall see that thwarting these sources of redundancy contributes to PETIoT slimness.}

Moreover, with VAPT over IoT devices as its goal, PETIoT focuses on the given device with its network environment, which often sees traffic with some supporting server or cloud as well as a companion software app to extend the minimal user interface of the device. Because of this, also the Installation and Command and Control steps of Lockheed Martin's CKC are unnecessary, upon the basis that no software needs to be maliciously installed and operated as in the case of intrusions into the device. \textcolor{black}{Also a reduction of these sources of redundancy will keep PETIoT slim.} 
We do not believe that the more recent and detailed KC listed above would add significant value to VAPT activities \textcolor{black}{over IoT devices}.

Finally, two more works are related. One, of 2018, tailors the KC model to multimedia service with its network environments~\cite{multimediakillChain}, but it seems to lack suitable practical demonstration. A more recent one, of 2020, derives a novel KC from a large number of IoT attacks~\cite{iotKillChain2020}; while this work is full of insights and relevant information, it seems specifically oriented at the creation of zombie devices towards DDoS.

\subsection{Security of IP cameras}
The related work concerning IP camera security hosts a number of items, because it has been shown that IP cameras may suffer multiple security issues, including multimedia security, network security and cloud security. However, only a few works are closely related to the contents of this article.

In 2015, Tekeoglu and Tosun looked into the security of cloud-based wireless IP cameras~\cite{cloudbasedIPcamerajpeg}. They studied the traffic generated by a wireless IP camera that is easy to set up for the average home user. Hence, they proved that an attacker can sniff the IP camera's network traffic and would be able to reconstruct the JPEG images from the data stream. Based on their information, their system has some limitations, e.g. their script only reconstructed 253 JPEG images over about 20 hours of video track. This work relates to ours in its focus on the device with its network environment, but its aims are more device-specific than ours of a general KC.

In 2017, Liranzo and Hayajneh presented an analysis of the main privacy and security issues affecting thousands of IP camera consumers globally~\cite{2017issueIPcamera}. They proposed a series of recommendations to help protect consumers' IoT devices in the intimacy of their houses, including security the (remote) connection between the smartphone running the video player and the camera. This is coherent with the observations and the findings reported in the present article.
Recently, Abdalla and Varol tackled an IP camera with similar aims to ours~\cite{abdalla}, but only contributed an assessment of basic vulnerabilities such as the use of default passwords. This and similar vulnerabilities would be captured during the central steps of PETIoT.

The Master thesis mentioned above~\cite{thesistapo} is of 2022 and targets the TAPO C200 too. It experiments with the Heartbleed vulnerability affecting the camera firmware up to version 1.0.10, which is prior to version 1.0.16 of our experiments. It also demonstrates an intrusion into the camera due to CVE-2021-4045 vulnerability, which affects firmware up to version 1.1.15, hence also the version of our experiments. However, that vulnerability was only published in March 2022, while our experiments date back to a year before, hence our automated scanners could not report it because they were lacking the relevant signature. With its focus on the device with its network environment, the sequel of this article demonstrates that the use of PETIoT unveils additional, unknown vulnerabilities.

\section{Defining PETIoT, a Cyber Kill Chain for the IoT}\label{sec:ourKC}

KCs can be understood as a comprehensive description of the various steps that must be followed to carry out some offensive activities. Our research started by considering existing KCs and continued by assessing how to leverage them towards answering our research questions. As a result, PETIoT was defined in a way to provide those answers and, therefore, it now guides the penetration tester's VAPT activities in the IoT domain.
In consequence:
\begin{itemize}

	\item \textbf{PETIoT spells out VAPT activities.} While we are used to KCs that only describe the steps of an intrusion or, separately, the steps to defend against that, PETIoT combines both for the first time simply because it aims at guiding VAPT activities. These always culminate with reports that not only describe prototypical exploitations of the vulnerabilities that were found, but also explain how to mitigate in practice if not entirely fix them.
	
	\item \textbf{PETIoT focuses on the IoT device.} Existing KCs bear some redundancy when applied to the IoT domain due to the fact that the target IoT device is made reachable to the tester and most of the needed tools are consolidated. Therefore, redundancy is appreciable in the early steps of Reconnaissance, which reduces to information gathering about the device, and Weaponization, which merely leverages a consolidated, basic toolkit.
	
	\item \textbf{PETIoT also focuses on the network environment.} While most KCs spell out intrusions into the target, 
	none pay sufficient attention to the environment, which is particularly relevant when the target is an IoT device, hence has traffic with supporting server or cloud and companion app. So, %
no offensive steps are needed after Exploitation because there is no software to operate remotely, while Actions on Objectives happens during Exploitation.

\end{itemize}

PETIoT prescribes the following 6 steps:

\begin{enumerate}
    \item \textbf{Experiment Setup.} \textcolor{black}{The tester is granted \textit{physical reach} to the IoT device to test, namely is given the device itself, or merely \textit{logical reach} to it, namely is provided with the device address such as an IP or a Bluetooth access. Reach is a plausible assumption because the target of the test is the device itself with its network environment, coherently with the scope of PETIoT. Appropriate setups follow, depending on the specific network interface to use to reach the device. The tester has to install the device in a realistic environment to enable herself to probe it by the relevant toolkit.}  The setup is termed \textit{non-invasive} when the device is installed and not tampered with. It is \textit{invasive} when the tester disassembles parts of the device with the aim of getting hold of relevant information that the device would never mean to transmit, such as a private key or a whitelist.
    \item \textbf{Information Gathering.}
    Similarly to other KCs, the tester attempts to acquire as much information as possible about the functioning of the device, also to unveil potentially hidden features. In a \textit{black-box session}, \textcolor{black}{that is, a testing session in which the tester has no information on the target environment at the beginning,} this is normally achieved by probing the device repeatedly through the setup prepared before. For example, crucial tests aim at understanding the use of cryptography, its type and working assumptions such as the protection of a cryptographic key into some Hardware Security Module. \textcolor{black}{
    A further capability, although exclusive for \textit{white-box sessions}, is that the tester may also assess source codes to derive additional information geared, as always, at identifying possible vulnerabilities. In fact, during a white-box session, the tester has usually access to all details regarding network, coding, policies, etc. The tester uses a standard toolkit (recalled below) to intercept, interpret and, if needed, decipher traffic in a passive (\textit{eavesdropping}) or in an active (\textit{man-in-the-middle)} fashion.}
    \item \textcolor{black}{\textbf{Traffic Analysis.}}
    It is always worthwhile to attempt to distill out the \textcolor{black}{entire network functioning of the device.}
    This effort may be interpreted as a reiteration of the previous step through structured experiments aimed at combining into the full picture the various insights gained about the network traffic. For example, a successful Traffic Analysis over the networking activity yields a sequence diagram detailing the exchanges that occur among all relevant participants. \textcolor{black}{However, it is clear that this step may not always succeed depending on the intricacies of the functioning of the target device as well as on its implementation of dedicated protection measures such as the use of randomised response mechanisms and of advanced cryptographic primitives.}
    \item \textbf{Vulnerability Assessment.}
    Once information is available about the target device, the \textcolor{black}{tester can move on to study, mechanically or manually, whether vulnerabilities exist. It means that,} while standard scanners such as OpenVAS, Nessus Pro and Nmap with scripting offer valid help, the tester may need to formulate parallel conjectures or develop special scripts depending on the type of target. For example, standard scanners may not come with software connectors for all possible target architectures. Vulnerabilities should be formulated as scenarios that would break at least the essential security properties of confidentiality, integrity and availability. The tester should be prepared to hunt for vulnerabilities at any architectural level, including hardware and software, and, in particular, \textcolor{black}{at any level of the communication protocols.}
    \item \textbf{Exploitation.}
    Not all vulnerabilities can be practically exploited, and this step aims at verifying whether that is the case. Exploitation perhaps is the step that requires most of the tester's creativity and stubbornness. It unfolds through repeated trials whereby the tester takes a rather empirical approach aimed at verifying whether a trial succeeds. For example, only  specific payloads may lead to success, while all others may not. Exploitation is also the core and main target of the entire VAPT activity. 
    The white-box approach or a successful Traffic Analysis following the black-box one may favour this step because the tester would be more easily driven through precise trials. \textcolor{black}{Following the stated focus on the IoT device environment, all offensive activities terminate in this step.}
    \item \textbf{Fixing.}
\textcolor{black}{Once the penetration tester successfully attacks the target device by the exploitation of a vulnerability, she must continue with an assessment on how to fix the attack.} Fixes could be of \textit{preventative nature} so that, once they get deployed over all similar devices on the large-scale production level, the vulnerability will no longer affect \textcolor{black}{neither the future devices, nor the updated ones}. When preventative fixes are not possible, the tester might revert to engineering fixes of \textit{notification nature}, \textcolor{black}{which aim at informing the device user when a scenario that may suffer the attack occurs --- in the hope that the user will know how to find secure ways to face that scenario. For example, if a browser cannot fully verify the certification of the displayed URL up to a trusted root authority,} it notifies its users of that because it was impossible to fully prevent the risk of a man-in-the-middle attack (since the certificate does not chain back to a valid root certificate).
    \end{enumerate}

\section{Basic toolkit and information}\label{sec:basics}

This Section recalls a well-known, basic toolkit in support of any VAPT activity, then continues by outlining the basic information that is useful to follow the sequel of the manuscript. 
We consider a basic toolkit to include the following software tools:

\begin{itemize}
    \item Ettercap~\cite{ettercap} is a suite of tools for Man-in-the-Middle attacks. It can be used to capture network packets, filter contents, dissecting protocols, etc. In this work, we will use it to eavesdrop between the Tapo application and the Tapo camera;
    \item Nessus~\cite{nessus} is a vulnerability scanner. It is generally used to assess the level of security of a node, or of an entire network, against known attacks and vulnerabilities. In this work, we will use it to perform a vulnerability assessment on the TAPO camera. 
    \item Nmap~\cite{nmap} is primarily an utility for network discovery. It can rapidly scan large network to discover running hosts and services. In this work, we will use it to gather information about the services exposed by the camera;
    \item SSL packet capture~\cite{sslpacket} is a strong debugging tool that allows to intercept encrypted traffic for subsequent analysis. In this work, we will use it to interpret SSL traffic to and from the Tapo application;
    \item IPtables~\cite{iptables} is an administration tool used for packet filtering and NAT. It allows an administrator to manage firewall rules in the Linux kernel, so that packets with certain characteristics are all handled correctly. In this work, we will use it to actively interpose between the Tapo application and the Tapo camera;
    \item Wireshark~\cite{wireshark} is network protocol analyser that allows to capture and analyse network packets, following streams such as TCP or HTTP, and decrypt protocols such as IPSec and TLS. In this work, we will use it to convert network traffic into playable video via the H264 extractor extension.
\end{itemize} 

Having seen the relevant tools in support of VAPT, further information is necessary to understand the activities conducted below to demonstrate PETIoT. As a start, the ability to appeal to the video streaming services may be gained by using the open source, multi-platform media player VLC~\cite{vlc} and the iSpy DVR~\cite{ispy}. Consolidated protocols exist to make the streaming work.
An essential protocol to integrate the camera with third-party infrastructures and systems, such as the Real Time Streaming Protocol (RTSP)~\cite{rtsp}.
IT is a network protocol used for video streaming and to orchestrate the exchange of media, such as audio and video, across the network. RTSP extends the RTP~\cite{rtp} and RTCP~\cite{rtcp} protocols by adding the necessary directives for streaming, such as:
\begin{itemize}
 \item Describe, used by the client to obtain information about the desired resource.
 \item Setup, used to specify how a single media stream is to be transported.
 \item Play/Pause, used to start or pause video playback.
 \item Record, used to store a stream.
\end{itemize}

Further relevant information comes from the Open Network Video Interface Forum (ONVIF), whose aim is to promote compatibility of video surveillance equipment so that devices made by different companies can interoperate. ONVIF provides an interface for communication with different devices through the use of various protocols. It also provides a few configuration profiles to make the best use of the different technical features.
In particular, profiles G, Q, S and T, are dedicated to video systems, however, the Tapo C200 is only compatible with Profile S~\cite{comp}.
Profile S is designed for IP-based video systems and involves the use of a compliant device (e.g. Tapo C200) and a compliant client (e.g. iSpy~\cite{ispy}) that can together configure, request and control video streaming. The most relevant features of Profile S concern user authentication, NTP support and H264~\cite{onvifs} audio and video streaming using RTSP.

One of the modern technologies to strengthen security protocols that are based on public-key cryptography is Certificate Pinning, which ``\textit{leverages knowledge of the pre-existing relationship between the user and an organization or service to help make better security related decisions}''~\cite{pinning}. For example, this is useful for a mobile application meant to connect to its supporting server or cloud. Because such a relationship is known since design time, the server's public key certificate is preinstalled in the application so that the latter will not follow up to encrypted traffic from a different peer even if that traffic is correctly certified.

Once an attack is found, measuring its severity is useful for several reasons, notably to promptly bootstrap the applicable processes during the incident management and response and, with less urgency, to inform the periodical cybersecurity and data protection risk assessment. A widely established approach to measure attack severity is the Common Vulnerability Scoring System (CVSS)~\cite{cvss}. While a few incarnations of CVSS exist, the widely established 3.0 version requires the tester to assign parameters Attack Vector (AV), Scope (S), Attack Complexity (AC), Confidentiality (C), Privileges Required (PR) \& Integrity (I), User Interaction (UI) and Availability (A). The system then returns a score ranging between 0 and 10, with the highest number indicating highest severity.

Finally, the Raspberry Pi 4 Model B device must be mentioned, a rather inexpensive device boasting a Broadcom BCM2711, Quad core Cortex-A72 (ARM v8) 64-bit SoC at 1.5GHz and several network interfaces. In particular, its IEEE 802.11ac Wi-Fi and Gigabit Ethernet features will be used below. When needed, its network interface card will be turned into an access point by means of Hostapd~\cite{hostapd}.

\section{Demonstrating PETIoT}\label{sec:demo}
PETIoT needs practical demonstration.  This is already available through the previous experiments that were instrumental to define and sharpen the very steps of our KC. For example, Experiment Setup and Information Gathering were proven essential for the first time over printers and phones~\cite{iotjournal_noblind}, while Traffic Analysis and Vulnerability Assessment turned out much larger in the application described below. Exploitation, in turn, was problematic over Amazon Echo Dot, a case in which the Fixing step inspired a new breed of preventative security measures that are still being tested~\cite{sergioasiaccs_noblind}.

\begin{figure}[ht]
 \centering
 \includegraphics[width=.35\linewidth]{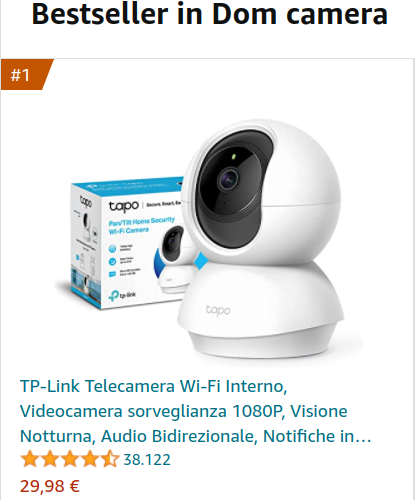}
 \caption{The Tapo C200 on Amazon Italy}
 \label{fig:tapoamazon}
\end{figure}

To offer a complete demonstration of PETIoT in the present manuscript, we chose a relevant IoT device that supports a paradigmatic application of the kill chain. 
Of course, the list of possible candidates is never ending, but we decided that the best balance between inexpensiveness and popularity lies in the area of small, portable cameras such as those used for home surveillance, often IP cameras~\cite{asmag}, and those typically used on vehicles, namely dash cameras~\cite{proclipdashcam}. 
With an estimate of around 770 million IP cameras used for CCTV in 2019~\cite{survcamera}, 
we decided to search through Amazon Italy and chose the no.1 best seller in category ``Dom camera'', which includes domestic cameras. As shown in Figure~\ref{fig:tapoamazon}, it is the Tapo C200 by TP-Link, which costs approximately 30€ and totals over 38.000 rather positive reviews at the time of this writing.

The C200 is an entry-level IP camera released in 2020. Its technical specifications support several functionalities, such as 
night vision, resolution up to 3MP, motion detection, acoustic alarm, data storage, voice control, use with third-party software and webcam mode.
We argue that such a popular target is ideal to demonstrate our kill chain, so the sequel of this article discusses how the main steps of PETIoT can be followed to conduct a VAPT session in practice on a real device. In this example, the tester is granted physical reach to the camera, conducts a black-box session of VAPT on it and concludes with a fix of preventative nature for one of the reported vulnerabilities.

\subsection{Experiment Setup}\label{sec:ES}
We begin by setting up our experiment over the C200. To operate the camera, 
a TP-Link account is required to enable access to the various cloud services available.
Access to the TP-Link account is via the Tapo app, which is available free of charge on Google Play for devices running Android 4.4+~\cite{tapoGPlay} and App Store~\cite{tapoApple} for those running IOS 9+. The application has a simple and intuitive user interface, from which it is possible to use and manage the devices of the Tapo family, including the Tapo C200. The range of services offered for the   C200 includes: remote access and camera control; sharing the camera with other accounts; synchronisation of settings; integration of the camera with smart home systems; receipt of motion notifications.

\begin{figure}[ht]
 \centering
 \includegraphics[width=.5\linewidth]{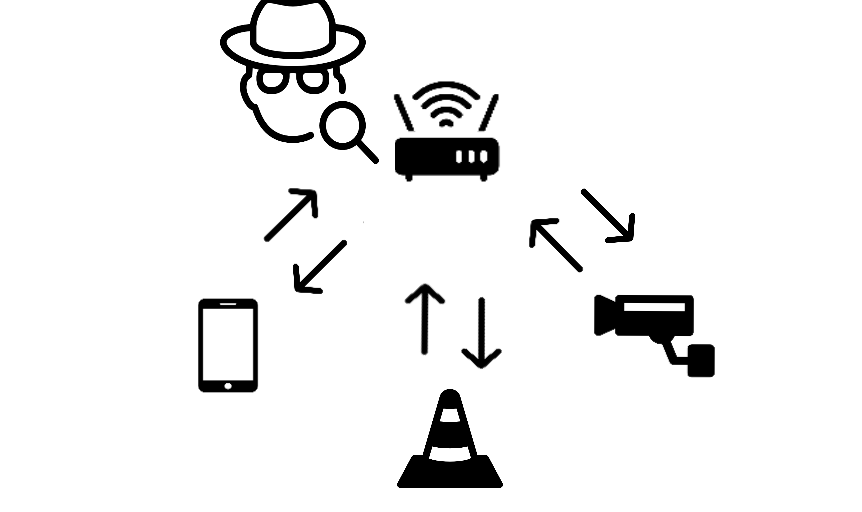}
 \caption{The testbed for the TAPO C200}
 \label{fig:testbed}
\end{figure}

Figure~\ref{fig:testbed} shows the testbed on which the experiments were conducted and how the various devices, including the attacker's machines are inter-connected via a Wi-Fi access point also operating as a switch. The testbed consists of:

\begin{itemize}
\item A Wi-Fi switch to which the various devices are connected.
\item The Tapo C200 IP camera.
\item A smartphone Oneplus 8T (Android 11, 8 Core, 8 GB RAM, 128 GB storage) running SSL Packet Capture and the Tapo application.
\item A Linux machine on which the attacker runs Ettercap and Wireshark.
\item A Linux machine on which the attacker runs the third-party players iSpy and VLC.
\end{itemize}

While the Tapo C200 works normally, our testbed enables us to monitor its network environment by means of the various tools with the aim of extracting useful information for operational analysis.

\subsection{Information Gathering}~\label{sec:IG}

Once the experiment is setup and ready, we move on to information gathering. 
The camera is now integrated with third-party infrastructures and systems, hence we can probe it through repeated sessions of experiments to gather our initial knowledge about its functioning. 
The basic information and toolkit outlined above guide us to conduct such experiments through this step.

Generally, an Nmap scan is the first tool that is used as it gathers information on open ports and exposed services. The Nmap manual offers a large range of scanning options, but simple ones are suitable in this case, mostly because stealthiness is not a requirement. Ettercap enables the tester's machine to interpose between the Tapo application and the Tapo camera, then Wireshark can be used to derive a dynamic view of the exposed services, namely the traffic they generate and receive. Appeals follow to SSL packet capture to interpret traffic oriented at the external TAPO server, and to Nessus to learn about known vulnerabilities that are in place.

In consequence of all the above trials, we learn about the services exposed by the camera and about the specific protocols that are used to make them available. Moreover, we find out that when the Tapo application and the Tapo camera are on different networks, a classic client-server architecture is used in which the Tapo Server is used to distribute information, such as the device configuration.
Conversely, when both are on the same network, no external servers are used to control the camera and access the video stream.

No known vulnerabilities are found by the automated Nessus scans conducted through this step. However, it is remarkable that three unknown ones are manually found later by combining our experience with the gathered information.

%Traffic Analysis
\subsection{Traffic Analysis}~\label{sec:RE}
Our work on this step tailors information gathering to an understanding of the precise message sequences that flow between the user, the app and the camera.
%
%It is important to note that the 
%The flow for control and editing differs substantially from that for video streaming. 
The Traffic Analysis effort is now partitioned depending on the software app that is used to stream from the camera, in particular on whether that software is the proprietary or third-party's.

\subsubsection{Using the Tapo C200 through proprietary software}
\begin{figure}[ht]
 \centering
 \includegraphics[width=.65\linewidth]{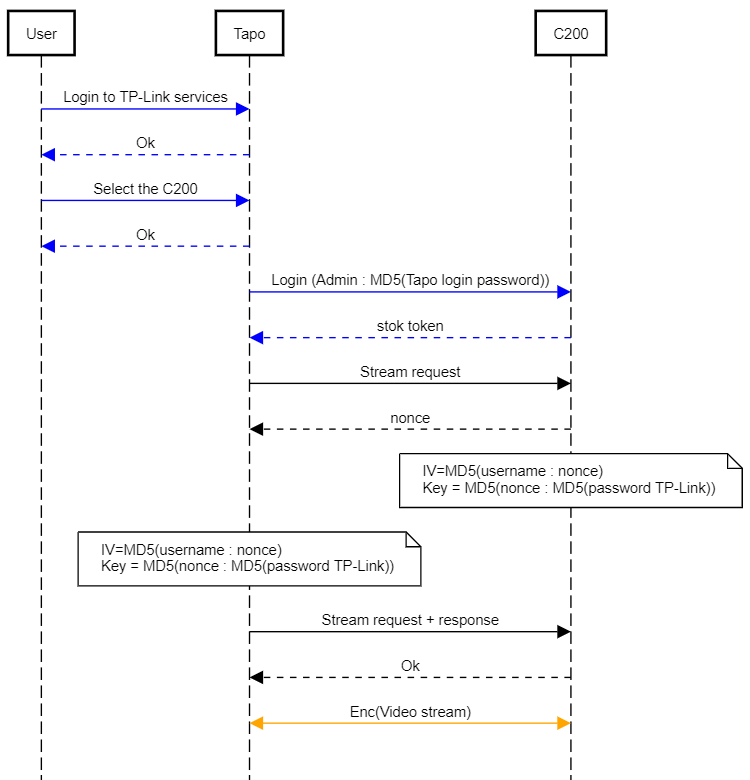}
 \begin{flushleft}
 \fcolorbox{black}{black}{\rule{0pt}{4pt}\rule{4pt}{0pt}}\quad Communication in plaintext\newline
 \fcolorbox{black}{blue}{\rule{0pt}{4pt}\rule{4pt}{0pt}}\quad Communication over TLS channel\newline
 \fcolorbox{black}{orange}{\rule{0pt}{4pt}\rule{4pt}{0pt}}\quad Stream video Encrypted with AES key
 \end{flushleft}
 \caption{Sequence diagram for using the IP camera with the proprietary Tapo application}
 \label{ctrltapo}
\end{figure}

The sequence diagram in Figure~\ref{ctrltapo} describes the various exchanges that allow the user to access the video stream through the Tapo application. After choosing one of the initialised devices, the Tapo application logs into the Tapo C200, obtaining the  \emph{stok} token, and this is needed to perform the control and modification operations.

%Stok
More precisely, the \emph{stok} is a token generated during authentication of the application to the camera. First, the Tapo C200 authenticates the user by a username and password, then assigns the \textit{stok} to the Tapo application. The purpose of the token is to create a session without having to re-authenticate the application every time the camera settings are changed. The various requests generated by the application carry the \emph{stok} obtained in the last authentication phase. In this way the Tapo C200 only accepts requests from authenticated users.

Following the user's request to access the video stream, the Tapo application and the Tapo C200 coordinate to serve that. In particular, the Tapo C200 sends to the application a \textit{nonce} that is necessary for the construction of a symmetric key.
At this point, both parties calculate the initialisation vector and an AES key in order to later encrypt the video stream using AES in CBC mode.
Once the key is calculated, the Tapo application authenticates itself to the Tapo C200 by providing a ``response'' tag. Finally, the Tapo C200, after verifying the validity of the response, starts the video streaming session by encrypting it with the agreed key. This completes the Traffic Analysis of the camera functioning when its proprietary app is used.

\subsubsection{Using the Tapo C200 through third-party software}
\begin{figure}[ht]
 \centering
 \includegraphics[width=.65\linewidth]{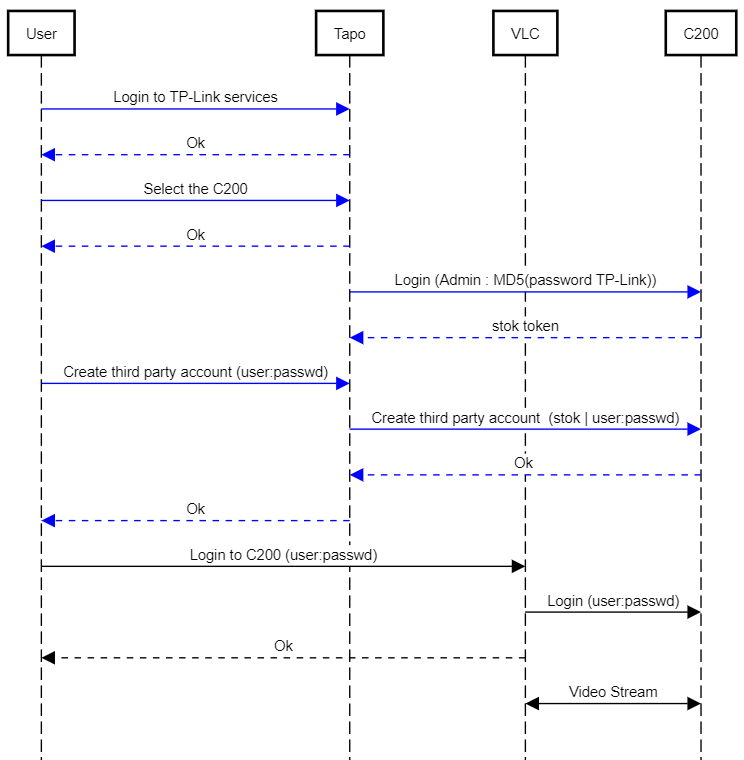}
 \begin{flushleft}
 \fcolorbox{black}{black}{\rule{0pt}{4pt}\rule{4pt}{0pt}}\quad Communication in plaintext\newline
 \fcolorbox{black}{blue}{\rule{0pt}{4pt}\rule{4pt}{0pt}}\quad Communication over TLS channel
 \end{flushleft}
 \caption{Sequence diagram for using the IP camera with third party software}
 \label{diagterz}
\end{figure}

Figure~\ref{diagterz} describes the outcome of our Traffic Analysis efforts targeting the camera use via third-party software. Using the Tapo application, we first need to access the device  and change its settings so that the camera can create a new user for third-party software.
After that, the Tapo C200 works with the software pair mentioned above, namely iSpy and VLC. These are used to carry out a login on the camera so that it initiates a free-to-air media streaming session.
In particular, the diagram shows that the \textit{stok} token is forwarded again to the camera, due to the necessary creation of the  dedicated user for the third-party app to access the camera. 

This diagram is derived by mounting a man-in-the-middle attack to the camera and analyse the traffic while the other machine running VLC and iSpy are accessing the camera and receiving the video stream. More precisely:

\begin{itemize}
 \item On VLC, select the ``network resources'' section of the player and enter:\\
 \texttt{rtsp://username:password@\\<tapoc200address>/stream/1}
 \item On iSpy, configure the ONVIF S profile to URI:\\ \texttt{http://username:password@\\<tapoc200address>/onvif/\\device\_service}
\end{itemize}

Since ONVIF's S profile does not add any functionality to the video stream, no differences emerged from the analysis of the traffic generated for the video stream of the two players. In particular, the packets analysed showed that, after access through the specific URI for ONVIF devices, iSpy uses RTSP in the exact way used by VLC. However, a remarkable difference is apparent by comparing the two diagrams discussed above, namely that, when third-party software engages with the camera, this starts an unencrypted video streaming session. It follows that such contents are intelligible to a man-in-the-middle attacker.

%VA
\subsection{Vulnerability assessment}~\label{sec:VA}
The sequence diagrams discussed above explain in great detail how the camera works, making a few significant features apparent. \textcolor{black}{In particular, we note that no traffic filtering mechanism is there, hence what follows.}

\begin{quote}
\emph{Vulnerability 1 -- \textcolor{black}{Improper neutralization of inbound packets allows complete Denial of Service.}} The camera uses no mechanisms to filter traffic, hence it would honour any message request as soon as possible, without checking the genuineness of the caller \textcolor{black}{and of the content}. It means that an attacker could maliciously attempt to flood it with traffic and make it unavailable.
\textcolor{black}{This vulnerability is a specific instance of weakness ``CWE-707: Improper Neutralization''~\cite{cwe707}.}
\end{quote}

\subsubsection{Using the Tapo C200 through proprietary software}

Our vulnerability assessment then continues by tackling motion detection. This important feature works as follows:

\begin{enumerate}
  \item The Tapo C200 detects some movement and sends a notification to the TP-Link server.
  \item The server receives the notification and sends an alert to all devices connected to the account associated with the camera.
  \item The application receives the message and generates a notification for the human.
\end{enumerate}

When the Tapo application and Tapo C200 are not on the same network, security in terms of the confidentiality of the video stream is entrusted to the channel itself on which the stream flows. Using SSL/TLS on port 443, all data exchanged between the Tapo C200, the TP-Link server and the Tapo application is properly encrypted.
In addition, the Tapo application and the Tapo C200 use certificate Pinning. It follows that, if information needs to be shared with the TP-Link server, no SSL/TLS session is established towards any entity other than a TP-Link server whose SSL/TLS certificate they already have by default. \textcolor{black}{We find no vulnerabilities about the implementation of these cryptographic protocols. }

\textcolor{black}{Our manual assessment} returns on the way motion detection is implemented, concluding that also the detection notifications are cryptographically protected hence should be unintelligible to the attacker. However, we observe a specific feature leading to a vulnerability.

\begin{quote}
\emph{\textcolor{black}{Vulnerability 2 -- Insufficient entropy in encrypted notifications allows breach of motion detection.}} The size of the motion detection notifications, despite the fact that these are encrypted, is always the same, precisely 523 bytes. We therefore report that, by using the Tapo C200 as an oracle, the attacker may be able to discern movement notifications based on the size of the messages, without intervening in the cryptographic scheme, which remains secure.
\textcolor{black}{This vulnerability is a specific instance of weakness ``CWE-331: Insufficient Entropy''~\cite{cwe331}.}
\end{quote}

\textcolor{black}{Following this style of Traffic Analysis, the attacker may deduce, for example, the precise times when someone is in the house that the camera monitors or when the house is empty. We shall see below in what scenario this vulnerability can be exploited. Additionally, by filtering and blocking only the messages carrying out a motion detection alert, the attacker could effectively deny the relevant notification to the legitimate user and convince the victim that no movement occurs in the area covered by the camera.}

\subsubsection{Using the Tapo C200 through third-party software}
The use of third-party players for video streaming is common because it is convenient for the end-user: it supports concurrent streaming from multiple cameras while the native app does not. With the increasing popularity of these devices, perhaps due to their low price and to the widespread need for surveillance, third-party players are becoming increasingly popular. Still, we found a feature that causes a vulnerability.

\begin{quote}
\emph{\textcolor{black}{Vulnerability 3 -- Cleartext transmission of video stream allows breach by unintended actors.}} Because the ONVIF profile S is limited to providing an interface for the use of RTSP, both streaming through ONVIF on port 2020 and RTSP on port 554 do not support encryption, hence there is no security measure to ensure confidentiality of the video stream.
In fact, after configuring the Tapo C200 to use these systems, whenever a request is made, the video stream is sent in the clear  by H264 encoding. This is clearly insecure, as the attacker could intercept the stream and decode it so that it would become fully intelligible. Also for this vulnerability, to what extent it can be exploited \textcolor{black}{will be seen below. This vulnerability is a specific instance of weakness ``CWE-319: Cleartext Transmission of Sensitive Information''~\cite{cwe319}.}
\end{quote}

%PT
\subsection{Exploitation}~\label{sec:E}

This Section reports on the exploitation experiments corresponding to the vulnerabilities discussed above. It is possible to exploit all three vulnerabilities reported above at least for an attacker who has access to the same network on which the TAPO C200 operates. This is the setup that is used through our exploitation efforts reported below.

\subsubsection{Denial of Service} \label{sec:camerados}
A Denial of Service (DoS) is a drastic malfunction of a device due to some form of resource exhaustion. Such attacks tend to undermine quality and, eventually, availability of a service by overloading it with an arbitrarily number of requests. 

\begin{figure}[ht]
 \centering
 \includegraphics[width=.4\linewidth]{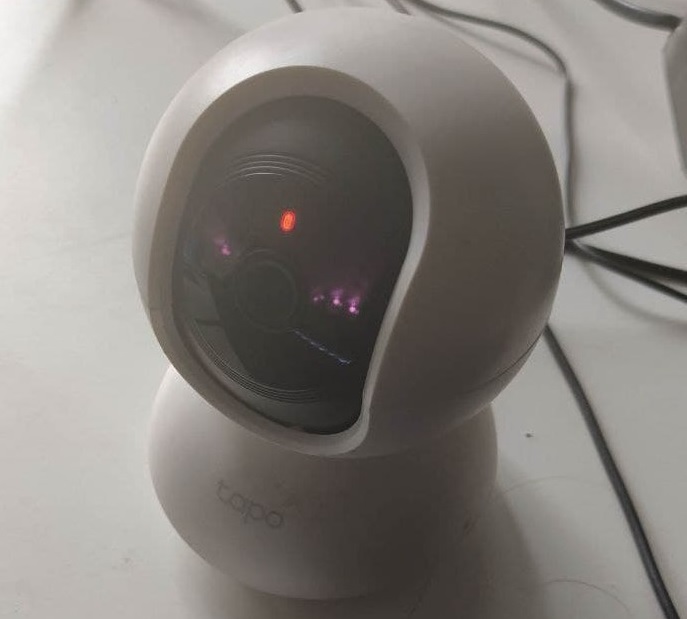}
 \caption{Crash of the Tapo C200 camera}
 \label{fig:crashed_camera}
\end{figure}

We succeeded in exploiting \textcolor{black}{the vulnerability that allows DoS on the Tapo C200 camera, that is, Vulnerability 1. In fact,} an intensive scan by Nessus makes the device crash, as it can be seen from the red led in Figure~\ref{fig:crashed_camera}. After the crash, the device reboots. 
As also occurred elsewhere, this may be due to ``\textit{insufficient memory specific to TCP/IP stack leading to a small amount of probes consuming it all and causing DoS}'' but we are ``\textit{unable to determine what the root cause is in this situation because we do not possess information about the TCP/IP implementation on the remote host}''~\cite{nessusfail}.

\begin{figure}[ht]
 \centering
 \includegraphics[width=.5\linewidth]{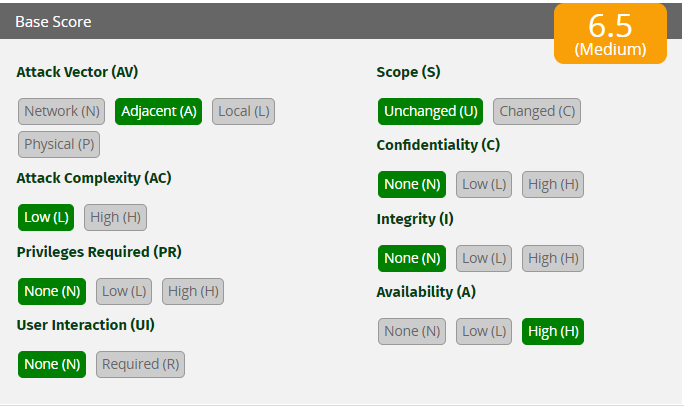}
 \caption{CVSS score calculated for Vulnerability 1 -- Denial of Service}
 \label{fig:DoSScore}
\end{figure}

Figure~\ref{fig:DoSScore} shows the CVSS score that we calculated for Vulnerability 1.
\textcolor{black}{As there is a complete loss of availability for every service offered by the camera, the Availability parameter is set to High, leading to the 6.5 CVSS score, which evaluates to Medium.}

\subsubsection{Breach of motion detection} \label{sec:motionoracle}

\begin{figure}[ht]
\centering
 \includegraphics[width=.9\linewidth]{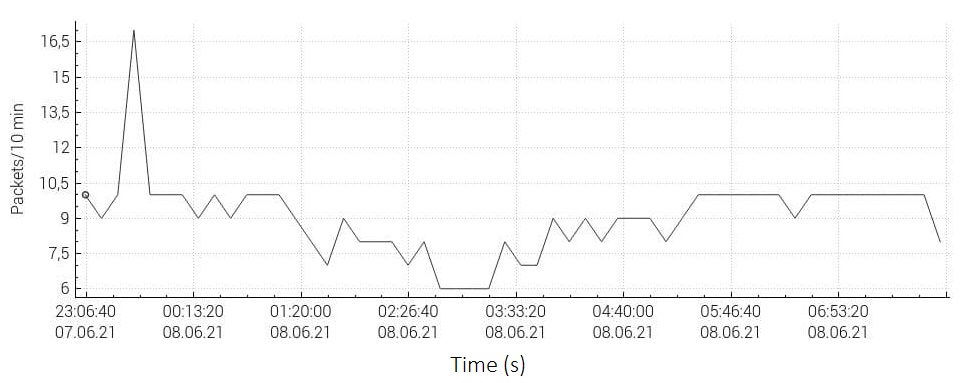}
 \caption{Packets carrying notifications of motion detection from an overnight capture}
 \label{graftraf}
\end{figure}

To exploit Vulnerability 2 and demonstrate that it is practically feasible,
we devise a special experiment consisting of two stages. A capture stage, namely to use the Tapo C200 to record a public place; a processing stage, namely to verify whether recorded events can be predicted correctly by intercepting \textit{encrypted} motion detection notifications. We conducted the capture part of the experiment by recording a suburban street of our hometown overnight back in June 2021. For the processing part, we intercepted packets using Ettercap then selected SSL/TLS packets of length 523 every 10 minutes. The outcome is represented in the chart in Figure~\ref{graftraf}. 

\begin{figure}[ht]
\centering
 \includegraphics[width=.5\linewidth]{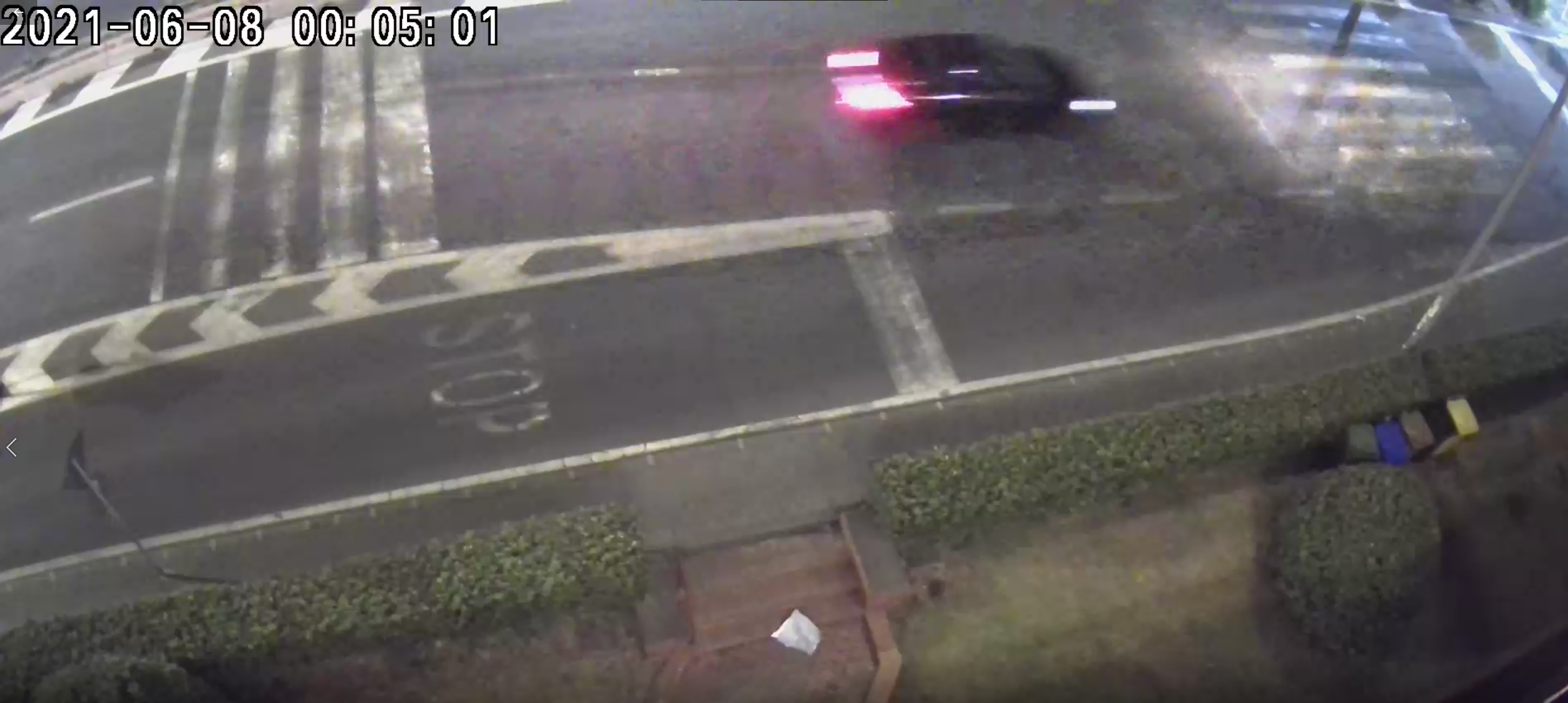}
 \caption{One of the frames at 00:05 recording a moving car}
 \label{immovi}
\end{figure}

It can be seen that starting at 23:00, the amount of packets begins to decrease until about 3:00. From 3:30 it is possible to see an increase in packets reaching a maximum after 5:00. We remark that the chart is solely built out of encrypted data. The trend of the curve is consistent with what could be expected from overnight traffic, but the experiment continued to confirm that by comparing the chart with the actual recorded video. As a result, the charted packets match recorded movements precisely.
For example, Figure~\ref{immovi} shows a frame taken at about 00:05, when a car can be seen in transit, while the chart shows 10 motion detection packets at the subsequent sampling time.

If on one hand our experiment can be interpreted as a confirmation of the reliability of the motion detection functionality, it becomes apparent, on the other hand, that such a function leaks relevant information due to the strictly deterministic use of encryption. As mentioned, an attacker could profitably take advantage of such an exploit. \textcolor{black}{If the camera is used to protect a household, the adversary could infer whether the legitimate user is at home or not, based on the captured motion detection notifications. Furthermore, Ettercap can be used to intercept motion detection messages and appropriate Iptables rules defined to prevent their delivery, thereby denying the motion detection service.}

\begin{figure}[ht]
\centering
 \includegraphics[width=.5\linewidth]{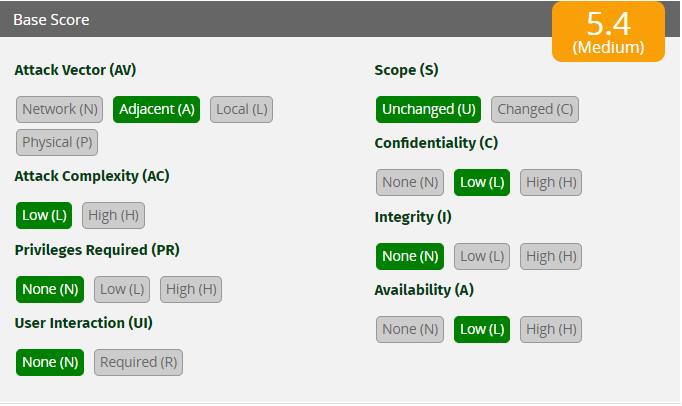}
 \caption{CVSS score calculated for Vulnerability \textcolor{black}{2} -- Breach of motion detection}
 \label{scoremo}
\end{figure}

Having verified how to exploit Vulnerability \textcolor{black}{2}, we are ready to score it by CVSS. Figure~\ref{scoremo} shows the resulting score, of 5.4, thus Medium relevance.

\subsubsection{Breach of video stream} \label{sec:privacyattack}
\begin{figure}[ht]
\centering
 \includegraphics[width=.5\linewidth]{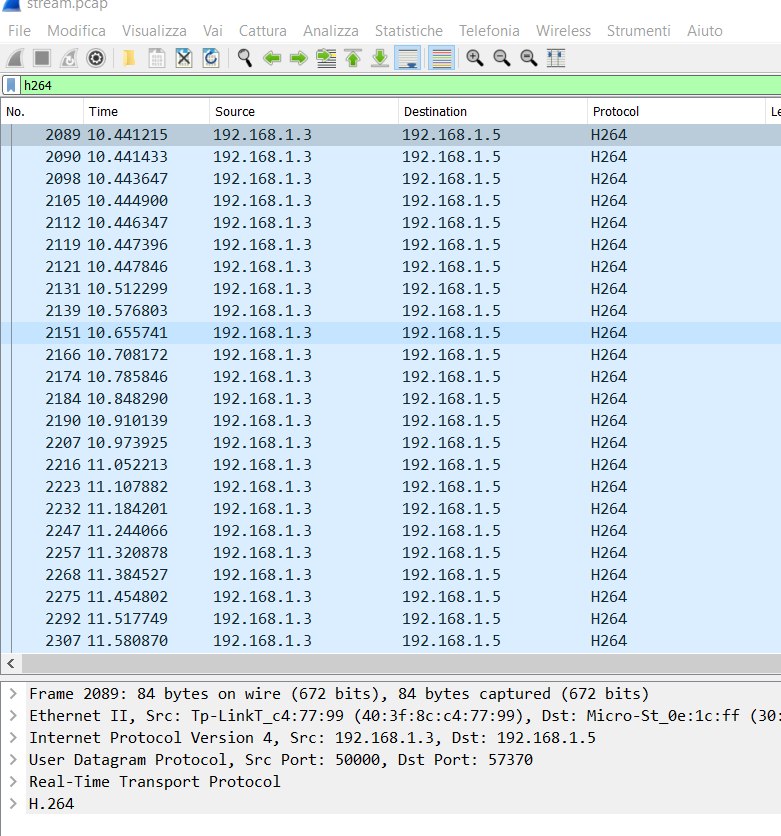}
 \caption{Example intercepted H264 frames}
 \label{extractor}
\end{figure}

To verify whether Vulnerability \textcolor{black}{3} can be exploited, we need to demonstrate how an attacker could obtain the video stream transmitted using third-party software. We conduct an experiment on the machine running the multimedia players by starting a multimedia session with the Tapo C200. At the same time, the other machine is acting as a MITM to intercept the traffic exchanged between the other two devices.
Traffic is first interpreted by using Wireshark. It is then possible to use the H264 extractor tool to reconstruct the entire video stream from its frames, which are shown in Figure~\ref{extractor}. %The output of the tool is a reproducible video of the intercepted session.
It is clear that lack of encryption makes the success of this exploit relatively easy to achieve.

\begin{figure}[ht]
\centering
 \includegraphics[width=.5\linewidth]{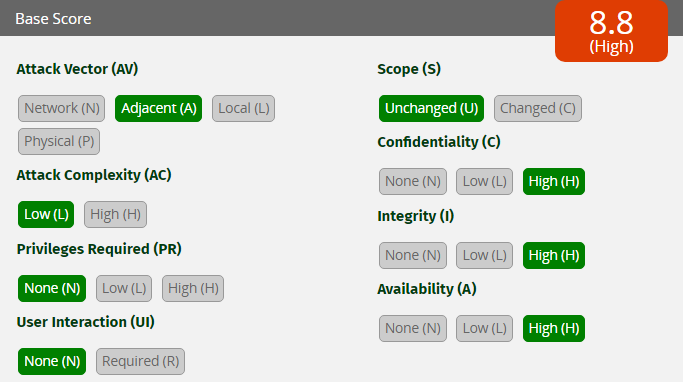}
 \caption{CVSS score calculated for Vulnerability \textcolor{black}{3} -- Breach of video stream}
 \label{fig:leakscore}
\end{figure}

Figure~\ref{fig:leakscore} shows the CVSS score we calculate for this vulnerability\textcolor{black}{: there is a High loss of Confidentiality as the adversary is able to read the whole video stream; Integrity loss is High as well, as the adversary could replace every H264 frame with an entirely different one, or could tamper with the sent frames; finally, Availability loss is High as the adversary could drop all H264 frames, thereby practically discarding the entire video stream. This leads to a CVSS score of 8.8 for this vulnerability, which evaluates to High severity.}

\subsection{Fixing}\label{sec:F}
The last step of PETIoT prescribes the definition of countermeasures for the vulnerabilities that may have been found and exploited.
In the particular case of the Tapo C200, the first vulnerability could be addressed by standard filtering strategies to mitigate denial-of-service attacks at kernel level and at firewall level. Vulnerability 2 could be addressed by randomising the size of the frame carrying the motion detection notification, and Vulnerability \textcolor{black}{3} by adding a cryptographic tunnel to the video stream. 

The implementation of such fixes clearly requires upgrades to the camera proprietary software, which we cannot do. However, we investigated various home-made setups as alternative fixes, and the sequel of this Section demonstrates them against Vulnerability \textcolor{black}{3}.
The idea behind our own solution to protect the video stream in the third party scenario is to use the Raspberry Pi 4 Model B as an access point of the Tapo C200 in order to modify the traffic in transit by applying a layer of encryption. \textcolor{black}{One Pi suffices to secure one camera but it must be remarked that this is only meant as a proof of concept of our fix, while the cryptographic tunnel that the Pi contributes to establish between the camera and the video player should be implemented by the vendor in the camera firmware instead.}

\subsubsection{Upgrading the sequence diagram}
\begin{figure}[ht]
\centering
 \includegraphics[width=.65\linewidth]{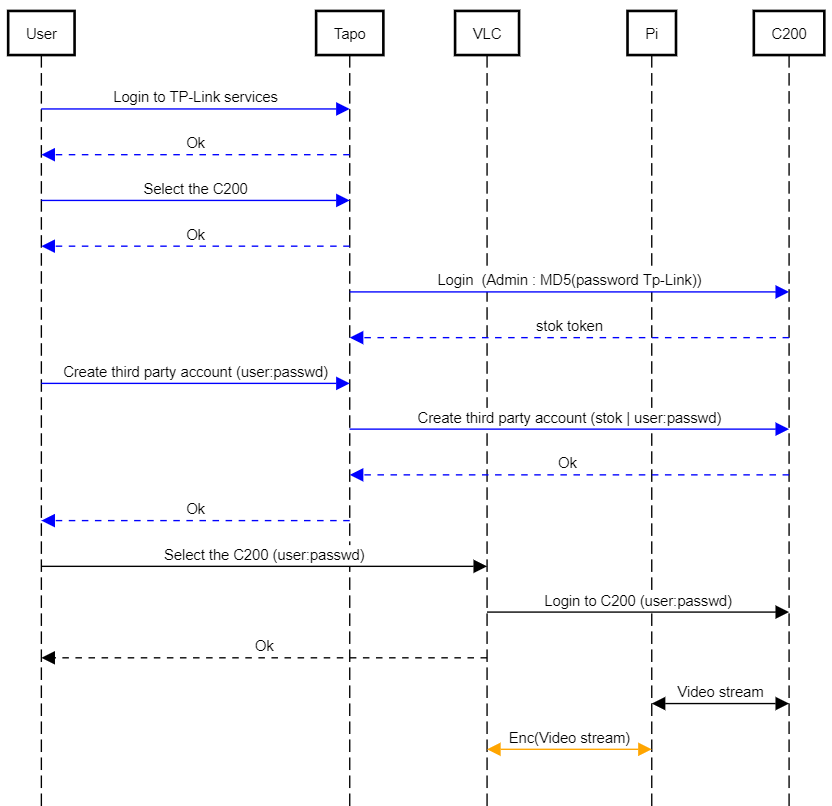}
 \begin{flushleft}
 \fcolorbox{black}{black}{\rule{0pt}{4pt}\rule{4pt}{0pt}}\quad Communication in plaintext\newline
 \fcolorbox{black}{blue}{\rule{0pt}{4pt}\rule{4pt}{0pt}}\quad Communication over TLS channel\newline
 \fcolorbox{black}{orange}{\rule{0pt}{4pt}\rule{4pt}{0pt}}\quad Video stream encrypted with symmetric AES key
 \end{flushleft}
 \caption{Sequence diagram for tunnelling the Tapo C200 video stream through a Raspberry Pi 4 Model B}
 \label{utipi}
\end{figure}

A sequence diagram representing the insertion of the Pi device when the Tapo C200 works with third-party software is shown in Figure~\ref{utipi}. 
It can be seen that the presence of the new device is totally transparent to the user, so that she can use the service exactly as she used to in the original scenario. %
The essential difference is that the Pi relays the cleartext video stream to the player but adds the desired encryption layer. It is clear that the new diagram makes sense against the same threat model considered before, which sees malicious activity within the same network that interconnects the devices. However, the upgraded diagram assumes that the attacker cannot sneak in through the connection between the camera and the Pi, an assumption that can be realistically enforced in practice by connecting solely the Pi and the camera on a dedicated SSID. %
\textcolor{black}{This is only a proof-of-concept of the fix, while the mentioned cryptographic tunnel should be implemented by the vendor in the camera firmware, so as to thwart the odds that the attacker interposes.}

\subsubsection{Upgrading the testbed}
\begin{figure}[ht]
\centering
 \includegraphics[width=.5\linewidth]{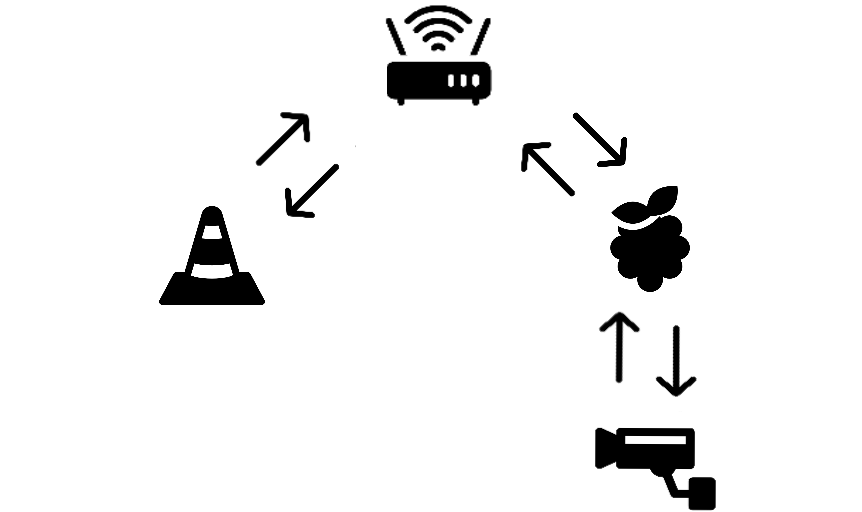}
 \caption{The testbed for the TAPO C200 with the Raspberry Pi 4 Model B}
 \label{testbed2}
\end{figure}

The Raspberry Pi 4 Model B has a network card that exposes a wired and a wireless interface. It was then possible to connect the device to the testbed switch by ethernet cable, while the Pi also exposes a separate Wi-Fi network to which the Tapo C200 can connect. Figure~\ref{testbed2} shows the resulting network  architecture.
More precisely, the Pi can be configured to act as an access point for the camera by Hostapd and some DHCP tweaks.

\subsubsection{Attempting exploitation of Vulnerability \textcolor{black}{3} again}
\begin{figure}[ht]
\centering
 \includegraphics[width=.5\linewidth]{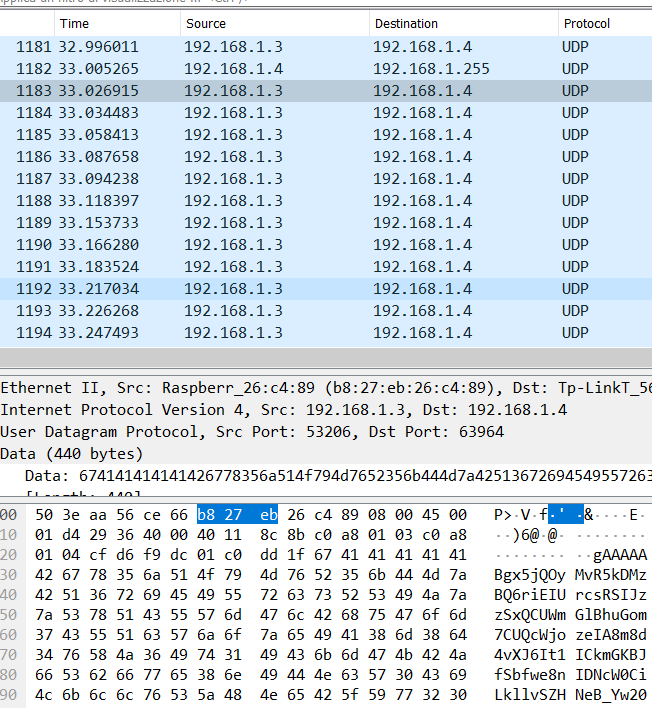} 
 \caption{A sample of encrypted network traffic}
 \label{trafficcif}
\end{figure}

In order to demonstrate the effectiveness of our home-made fix, the same experiment to exploit Vulnerability \textcolor{black}{3} was carried out again, but this time the execution of the H264 had no luck.  Figure~\ref{trafficcif} shows that the intercepted traffic is unintelligible, hence the execution of the H264 extractor tool could not produce valid results.

\subsubsection{A glimpse at the implementation of encryption}
A fundamental prerequisite was to configure IPtables on the Raspberry Pi with the policy in Listing~\ref{code:IPtablesEnc}:

\begin{lstlisting}[caption={Configuration of IPtables for encryption},  label={code:IPtablesEnc}]
iptables - A FORWARD - m 
--physdev-is-in wlan0 -p udp
-s TAPOC200_ADDRESS_IPV4
-j NFQUEUE --queue num 1

iptables -A INPUT-A FORWARD -m
--physdev-is-in wlan0 -f -j DROP
\end{lstlisting}

The first rule makes it possible to intercept packets in transit (FORWARD chain) from the network interface (-m --physdev-is-in wlan0) transported with UDP (-p udp) coming from the Tapo C200 (-s TAPOC200\_ADDRESS\_IPV4) and then to send them on queue 1.
After that, the second rule eliminates all possible fragments of the packets in transit, in order to avoid sending partial packets.

Once the firewall queues up the packets as discussed, the Python script shown in Listing~\ref{code:encryptscript}, encrypts the packets in the queue. The script retrieves packets from the queue, extracts the payload (by skipping 28 bytes of headers), encrypts it and sends it back to the recipient's address.

\begin{lstlisting}[caption={Encryption script}, label={code:encryptscript}]
def encrypt(packet):
  cipher_suite = Fernet(key)
  encoded_text = cipher_suite.encrypt
  (packet.get_payload()[28:])
 
  pkt = IP(packet.get_payload())
  UDP_IP = pkt[IP].dst
  UDP_PORT = pkt[UDP].dport
  MESSAGE = encoded_text
  
  sock = socket.socket(socket.AF_INET,
  socket.SOCK_DGRAM)
  sock.sendto(MESSAGE, 
  (UDP_IP, UDP_PORT))

  packet.drop()
  
\end{lstlisting}

\subsubsection{A glimpse at the implementation of decryption}
A configuration of IPtables was also necessary for the client machine running third-party video player in order to enable it to decrypt the encrypted stream originating with the Pi. Listing~\ref{code:IPtablesDec} shows the policy.

\begin{lstlisting}[caption={Configuration of IPtables for decryption}, label={code:IPtablesDec}]
iptables -A INPUT -p udp
-s RASPBERRY_ADDRESS_IPV4
-j NFQUEUE --queue num 2 
\end{lstlisting}

This policy intercepts incoming packets (INPUT chain) transported with UDP (-p udp) coming from the Pi (-s RASPBERRY \- ADDRESS \_ IPV4) and then sends them on queue 2.

The corresponding script for decrypting the packets is shown in Listing~\ref{code:decryptscript}.
It fetches packets from the queue, extracts the payload and decrypts it before accepting it.

\begin{lstlisting}[caption={Decryption script}, label={code:decryptscript}]
def decrypt(packet):
  cipher_suite = Fernet(key)
  decoded_text = cipher_suite.decrypt
  (packet.get_payload()[28:])
 
  pkt = IP(packet.get_payload())
  UDP_IP = pkt[IP].dst
  UDP_PORT = pkt[UDP].dport
  MESSAGE = decoded_text
  
  sock = socket.socket(socket.AF_INET,
  socket.SOCK_DGRAM)
  sock.sendto(MESSAGE,
  (UDP_IP, UDP_PORT))
  
  packet.drop()

\end{lstlisting}

\section{Lessons learned and recommendations}\label{sec:lessons}
The research and developments discussed above taught us that the notion of a Kill Chain is used rather broadly in the ethical hacking community. Therefore, this article aimed, first and foremost, at preciseness. The first limitation arising from the state of the art was the separation between KCs for attack or for defence, whereas both are indispensable during VAPT activities, hence PETIoT combines them. Moreover, we learned and experienced that each KC comes with a precise focus, which, by necessity arising from the (im)maturity of the state of the art, is not always taken strictly by practitioners. For example, Lockheed Martin's CKC concerns intrusions into a target device, other KCs detail the intrusion process further, while others still concern operations that are internal to the target. Then, we saw how steps that are established within other KCs become redundant in PETIoT due to the focus on the IoT domain, or unnecessary due to the focus on the device environment. 

What is noted above allowed PETIoT to stay very slim hence, arguably, easily applicable in practice, so that it is the recommended choice for the penetration tester who is tasked with an IoT device, providing the most effective steps to focus on the device environment.
Even as a result of this recommendation, which has been demonstrated throughout the present article, it is firm that the state of the art remains useful. For example, a simplification of Lockheed Martin's CKC as explained above is recommended to guide actual intrusion activities in an IoT device because the KC is general about intrusions into \textit{any} device. Notably, a recent Master's thesis~\cite{thesistapo} may be interpreted in support of this recommendation despite the fact that the thesis itself does not make it explicit that it is, in fact, applying a KC. Conversely, PETIoT could be applied to any device beyond the IoT to instruct VAPT activities focused on the environment of the device.

The very demonstrator that was tackled, the TAPO C200, also taught us general lessons. Because flooding this device (and similar ones, as noted over other devices discussed elsewhere) is rather simple, not just household chores but also sensitive monitoring such as over newborns or house perimeters are at risk. Therefore, it is recommended that this and every IoT device implements appropriate traffic filtering measures. Also, a robber could profitably exploit the breach of motion detection to infer when a target household is very likely to be unmanned. The consequent recommendation is to randomise the motion detection signal on this and similar devices to minimise its predictability. Finally, the video stream breach is clearly intrusive, hence a natural recommendation is that no unencrypted stream ever leaves any IP camera, independently of the use case. The camera vendor acknowledged all three vulnerabilities through the responsible disclosure process that we followed with them, so that the current firmware version of the TAPO C200 fixes all vulnerabilities discussed in this article.

\section{Conclusions}\label{sec:concl}
This article aimed at facilitating the adoption of the KC approach for the sake of ethical hacking in the IoT application domain. By defining the PETIoT KC for IoT devices, the article showed how penetration testers can follow a KC during VAPT sessions.
PETIoT is slim and readily applicable. Its focus is on how the tester can violate a device she \textcolor{black}{can reach, either physically or logically}. Our KC insists on a fixing step for devising preventative or notification measures and thwart the vulnerabilities that may have been found and exploited. 

\textcolor{black}{In consequence, our research questions can be answered by noting that, in general, there do exist approaches that can be leveraged towards VAPT in the IoT domain albeit with appreciable tailoring and that, in particular, PETIoT can be applied beside a simplified version of Lockheed Martin's CKC.}

To demonstrate PETIoT, we followed it to engage into a VAPT session on the best-selling IP camera on Amazon Italy, the TAPO C200 by TP-Link. \textcolor{black}{Under specific circumstances, three vulnerabilities arose, one in terms of denial of service, the other two in terms of privacy of the camera users and overall security of the services that the camera exposes to them. It is found that the tester can, respectively, understand motion detection signals from cryptographically protected notifications, and breach the captured video when it is streamed unencrypted towards a third-party player. With a CVSS score of  8.8, the last vulnerability, in particular, is of High severity, and a proof-of-concept fix for it was provided. Responsible disclosure with the vendor led to fixing all reported vulnerabilities in the latest firmware released for the camera.}

PETIoT has been gestated out of years of experiments with several types of IoT devices. We have been evaluating its benefits empirically over the many devices our research group has been besieging. The outcomes have always been very positive, confirming that PETIoT effectively drives the testers through the operations, saving time, %A
\textcolor{black}{although any KC may require several sub-steps such as scripting, fuzzing, probing, physical tampering, etc, which depend on the precise architecture and functions of the target device as well as on the tester's skillset.}
Future work includes further applications of our KC to more devices, and this will be facilitated by the popularity of the techniques experienced over the TAPO C200, including ONVIF profiles and RTSP. It seems fair to conclude that our demonstration of PETIoT gained us an in-depth understanding of the target camera from a cybersecurity and privacy standpoint, \textcolor{black}{hence it is well worth employing on more IoT devices}.

\section{Acknowledgements}
We are grateful to Angelo Zinna for contributing to the experiments. 
This research was supported by ``Linea di intervento 2, progetto MEGABIT'' of funding programme ``PIAno di inCEntivi per la RIcerca di Ateneo 2020/2022 (PIACERI)'' of University of Catania, and open access was supported by ``Linea di Intervento 4'' of the same funding programme.

\end{document}